\documentclass[acmsmall]{acmart}
\AtBeginDocument{%
  }

\setcopyright{acmlicensed}
\copyrightyear{2025}
\acmYear{2025}
\acmDOI{XXXXXXX.XXXXXXX}
\acmConference[XXX]{Make sure to enter the correct
conference title from your rights confirmation email}{XXX,
XX}{XXX}
\acmISBN{978-1-4503-XXXX-X/2018/06}

\usepackage{balance}

\begin{document}

\title{Towards Advancing Code Generation with Large Language Models: A Research Roadmap}
\author{Haolin Jin}
\email{hjin3177@uni.sydney.edu.au}
\affiliation{%
  \institution{University of Sydney}
  \city{Sydney}
  \state{NSW}
  \country{Australia}
}
\author{Huaming Chen}
\email{huaming.chen@sydney.edu.au}
\affiliation{%
  \institution{University of Sydney}
  \city{Sydney}
  \state{NSW}
  \country{Australia}
}
\author{Qinghua Lu}
\email{Qinghua.Lu@data61.csiro.au}
\affiliation{
  \institution{CSIRO's Data61}
  \city{Sydney}
  \state{NSW}
  \country{Australia}
}
\author{Liming Zhu}
\email{liming.zhu@data61.csiro.au}
\affiliation{
  \institution{CSIRO's Data61}
  \city{Sydney}
  \state{NSW}
  \country{Australia}
}

\begin{abstract}
Recently, we have witnessed the rapid development of large language models, which have demonstrated excellent capabilities in the downstream task of code generation. However, despite their potential, LLM-based code generation still faces numerous technical and evaluation challenges, particularly when embedded in real-world development. In this paper, we present our vision for current research directions, and provide an in-depth analysis of existing studies on this task. We propose a six-layer vision framework that categorizes code generation process into distinct phases, namely Input Phase, Orchestration Phase, Development Phase, and Validation Phase. Additionally, we outline our vision workflow, which reflects on the currently prevalent frameworks. We systematically analyse the challenges faced by large language models, including those LLM-based agent frameworks, in code generation tasks. With these, we offer various perspectives and actionable recommendations in this area. Our aim is to provide guidelines for improving the reliability, robustness and usability of LLM-based code generation systems. Ultimately, this work seeks to address persistent challenges and to provide practical suggestions for a more pragmatic LLM-based solution for future code generation endeavors.
\end{abstract}

\ccsdesc[500]{Software and its engineering}

\keywords{Large Language Models, Code Generation, LLM-based Agent}

\maketitle

\section{Introduction}
\begin{figure*}
    \centering
    \includegraphics[width=\linewidth]{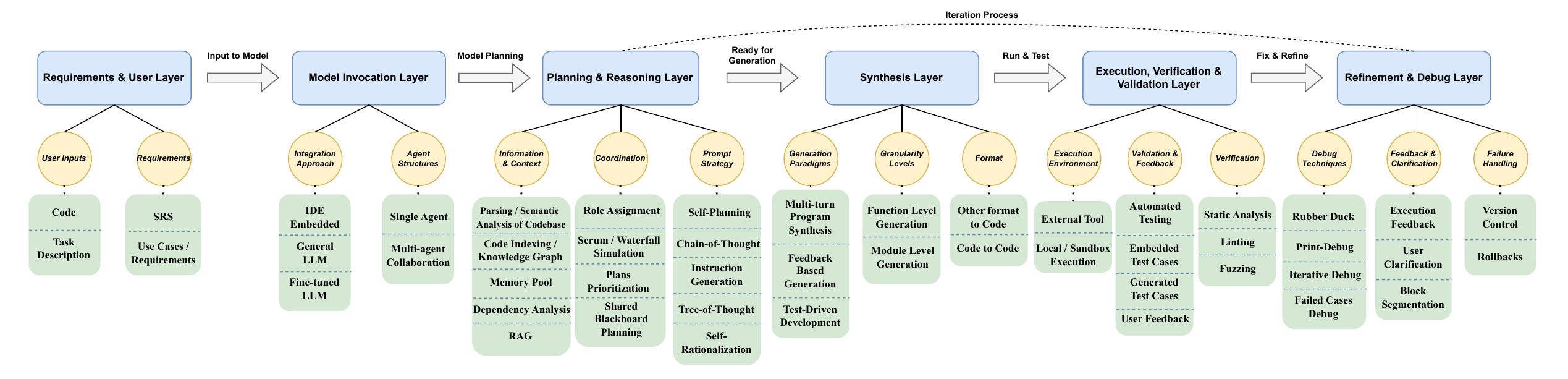}
    \caption{Proposed Six-Layer Architecture for LLM-Based Code Generation.}
    \label{fig:framework}
\end{figure*}
Code generation, also known as program synthesis, aims to automatically generate code based on specified inputs \cite{austin2021program}. In recent years, Large Language Models (LLMs) emerge as a prominent solution to generate code from natural language descriptions, saving time for developers and serving as benchmarks for evaluation \cite{chen2021evaluating,nijkamp2022codegen}. Evaluation typically involves generating code from descriptions of natural language and validating it against the test sets \cite{ni2024l2ceval,austin2021program,hendrycks2021measuring,lu2021codexglue}. Other key tasks include code translation \cite{ruder2019unsupervised,zhong2024ldb} and comprehension \cite{feng2020codebert}. However, the evaluation process may introduce bias since it heavily relies on experienced professionals. In the meantime, specialized models like CodeGen and DeepSeek-Coder have been developed for tasks such as code completion, optimization even debugging and reviews. These models demonstrate the growing potential of LLMs in automating programming tasks.

Despite the wide usage for basic coding, professional programmers prioritize the accuracy and robustness of the generated code. It is identified that inherent shortcomings in LLMs, such as limited contextual understanding \cite{kaddour2023challenges,hadi2023survey} and hallucinations \cite{ji2023survey}, may lead to contents with semantic and syntactic errors. To address these challenges, LLM-based agent frameworks are presented for self-refinement and enhancing the model's understanding of tasks and the accuracy of responses with prompt engineering. While they have shown significant improvements over baselines, they still face limitations and challenges across multiple studies.


Previous works explore the potential of LLMs for code generation and introduce benchmarks such as HumanEval \cite{chen2021evaluating} and MBPP \cite{austin2021program}. The reasoning capabilities of LLMs have also inspired various survey works, highlighting their potential to automate mundane coding tasks, allowing programmers to focus on higher-level activities \cite{chang2024survey,xi2023rise,jiang2024survey}. However, these works place strong emphasis on model performance evaluation, offering limited insights into framework-level implementation and analysis. For instance, recent attention has been given to LLM-based agent frameworks \cite{xi2023rise,jin2024llms}, yet the construction and implementation of these frameworks have not been thoroughly examined.

In this paper, we present our vision and reflections on the application of LLMs in code generation, featuring their current advancements while sharing our perspectives on future researches. We also explore the challenges in code generation from both technical and evaluation viewpoints, offering our proposals as potential solutions. Although extensive research has focused on LLMs' downstream tasks, particularly code generation, focusing efforts on fundamental objectives for pragmatic solutions over merely achieving higher benchmarks is crucial. Through this work, we aim to provide the community with a comprehensive overview of the current trends and challenges in LLM-based code generation, fostering deeper understanding and more impactful future research.

\section{Vision for LLM-Based Code Generation}
LLMs' capabilities have evolved from basic sentiment analysis to creative tasks~\cite{chen2024large}, with significant improvements from early models like GPT-3.5 to the more advanced GPT-4, featuring a larger parameter scale, improved learning schemes, and deeper alignment with human feedback~\cite{kalyan2024survey,wu2023next}. However, existing works often highlight repetitive architectures and workflows. To investigate this, we propose a six-layer vision framework to distill core components and streamline the foundation for more efficient and practical solutions.
\subsection{Conceptual Framework} \label{sec:framework}
\textbf{\textit{Layer 1-3}} \ \ Figure~\ref{fig:framework} presents a six-layer collective architecture for how LLMs handle code generation task in current works. The first layer captures user input or the task requirements; with common use cases highlighted in green. This flows into the second Model Invocation Layer, which encompasses different modeling approaches. For example, `Cursor'\footnote{\href{https://www.cursor.com/}{https://www.cursor.com/}} act as an IDE using LLM APIs for code generation and basic chat functions, while models could be further fine-tuned specifically for code generation. Following that, the agent or model employs prompt engineering for planning and self-reasoning. We observe most LLM-based code generation studies include this step, frequently employing role assignment and task decomposition, as seen in MetaGPT, ChatDev's Chat Chain~\cite{hong2023metagpt,qian2024chatdev}.
Context Understanding involves techniques like code parsing and semantic analysis to build an internal representation of the project. Methods such as knowledge graphs and dependency analysis help map relationships among files, libraries, and APIs. Retrieval-Augmented Generation (RAG) \cite{lewis2020retrieval} further enhance the model's knowledge base with external documentation and repositories.

\textbf{\textit{Layer 4-5}} \ \ The fourth layer, the synthesis layer, involves code generation after planning and reasoning. Similar techniques are included in many agent frameworks to increase code accuracy, such as feeding feedback into the model's analysis to reduce computational overhead and ensure adherence to guidelines \cite{chen2023teaching}. Additionally, multi-turn program synthesis, as in CodeGen \cite{nijkamp2022codegen}, is another typical paradigm, with formats for synthesis including natural language-to-code and code-to-code translation. The fifth layer involves running the generated code. Many frameworks utilize external tools \cite{zhang2024experimenting}, while some studies advocate manual validation \cite{zhang2023self}. Validation can be categorized into two groups: using built-in test sets from benchmark itself, or test sets generated by LLMs \cite{huang2023codecot}.

\textbf{\textit{Layer 6}} \ \ The final layer, Refinement \& Debug, includes three key modules. A common method is iterative debugging, where the model performs self-refinement based on prior runtime results, often focusing on erroneous test cases for separate analysis \cite{chen2023teaching}. Another feedback mechanism is user clarification, addressing the ambiguous task descriptions by requesting user input for clearer instructions \cite{mu2023clarifygpt}. As for failure handling, which is not yet widely used, it typically involves storing code snapshots in a registry \cite{holt2023l2mac}. If errors occur, the system can revert to a previous version, a workflow reminiscent of daily programming practices. If the result remains unsatisfactory after this layer, a common approach is to return to the planning and reasoning phase and repeat a similar process until the code passes tests or the maximum iterations limit is reached.
\subsection{Our Vision} \label{subsec:vision}
\begin{figure*}
    \centering
    \includegraphics[width=\linewidth]{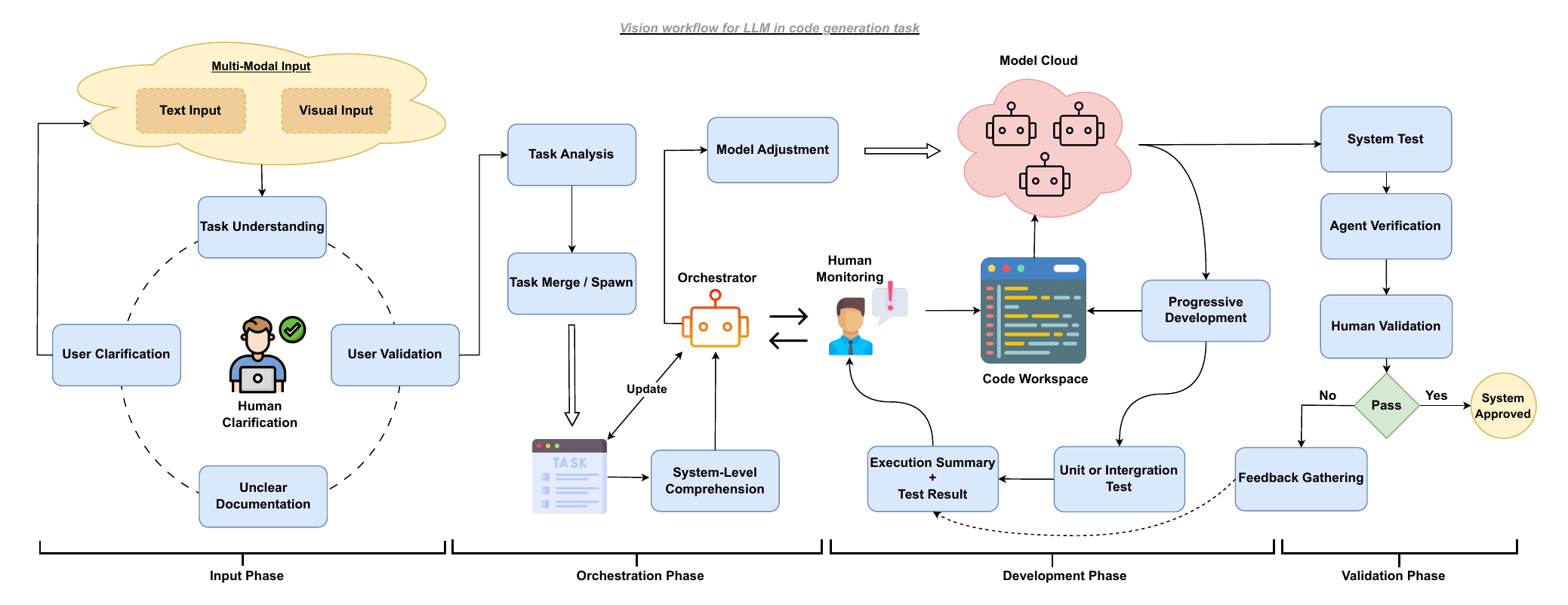}
    \caption{Our vision workflow for LLM-Based Code Generation.}
    \label{fig:vision}
\end{figure*}
In Section~\ref{sec:framework}, we discuss our perspective on how LLMs approach code generation, providing a detailed analysis of the overall architecture. Figure~\ref{fig:vision} illustrates our proposed vision workflow, structured into four phases: Input, Orchestration, Development, and Validation. Unlike the six-layer architecture, our vision does not aim to construct a concrete framework. Instead, it synthesizes insights from existing works to highlight how LLMs have fundamentally transformed software development, providing a comprehensive workflow that serves as a clear roadmap for future research.

\textbf{\textit{Input Phase}} \ \ Currently, LLMs in code generation predominantly follow a `prompt in, code out' approach. Our vision expands this with multi-modal inputs, such as flowcharts as supplementary diagrams, to better reflect real-world software development and reduce the communication overhead~\cite{huang2023chatgpt}. However, LLMs often generate erroneous or non-functional code due to lack of clear information~\cite{ouyang2024empirical,hendrycks2021measuring,yang2024robustnesssecurityprivacyexplainability}. Discrepancies in task descriptions can severely hinder the model's comprehension, resulting in outputs that fail to meet user requirements.
To address this, our vision incorporates a \underline{clarity check} process. Our approach advocates for `frequent human-model interaction', allowing users to verify LLMs' understanding of tasks and clarify ambiguities. Similar to ClarifyGPT \cite{mu2023clarifygpt}, the goal is to reduce hallucinations or speculative outputs from LLMs, and improve LLMs reliability for code generation.

\textbf{\textit{Orchestration Phase}} \ \ The concept of `orchestration', originally from cloud computing \cite{weerasiri2017taxonomy}, is now widely used in multi-LLM agent frameworks for collaborative operations \cite{xi2023rise}. Our vision introduces \underline{Dynamic Task Creation}, enabling on-demand task generation and iterative refinement during development, leveraging execution results and human feedback. During the orchestration phase, the orchestrator LLM performs \underline{system-level comprehension} of the current task list to dynamically adjust the agents number based on task complexity. This step enables the agent to achieve a clearer understanding of the overarching task and workflow. Once the orchestrator attains a comprehensive understanding of the development tasks, it executes model adjustment to generate multiple agents to support subsequent development processes. These agents are then stored in a model cloud for further utilization.

\textbf{\textit{Development Phase}} \ \ Traditional LLM-based code generation often focuses on module or function level outputs, and simulate workflows like Scrum or Test-Driven Development (TDD) \cite{lin2024soen}. However, for multi-agent frameworks, it can incur heavy overhead due to the crowd collaboration of agents. Also it lacks transparency since their intermediate processes are typically opaque (black-box), increasing the risk of introducing errors or security flaws \cite{qian2024chatdev}. Moreover, these methods lack the explainability that many developers and stakeholders require, as it can be difficult to trace how decisions are made or how code components evolve.

Our vision workflow promotes frequent human interaction over autonomous generation. Since the ultimate goal of any software product is to serve human needs, consistent and deliberate user scrutiny remains indispensable. 
In our workflow, users (or development team) actively monitor the LLM's incremental coding process—testing and reviewing each module or function upon completion. This approach mirrors real-world practices, such as front-end developers visually inspecting features as they are written. Ongoing \underline{workspace monitoring} enables users to collaborate with the orchestrator to address errors or unexpected outcomes, refining tasks and agents mid-development. The developer-in-the-loop paradigm combines human oversight with LLM capabilities, reducing black-box risks, enhancing explainability, and ensuring robust and efficient code generation. 

\textbf{\textit{Validation Phase}} \ \ In our vision, this phase emphasizes \underline{system-} \underline{level testing} and \underline{human validation}, as debugging and iterative fixes occur during the Development Phase. Once the code reaches a stable state, the orchestrator conducts system tests, verifying all components, including newly generated modules and integrated code, interoperate as intended under real or simulated production conditions. Successful system tests lead to human validation, where developers and users evaluate functionality, usability, and alignment with project goals. This step ensures that the product meets real-world expectations—covering edge cases, performance targets, or business requirements that might fall outside automated checks. Should any discrepancies arise, the workflow routes back to the Development Phase, allowing the team to integrate the system test results and human feedback into further refinements or expansions of the code. The orchestrator then reassigns tasks based on these new insights, enabling a cyclical process of improvement.

\section{Challenges and Reflections} \label{sec:challenge}
In this section, we primarily explore common challenges prevalent across different current studies. Additionally, we provide suggestions and insights on how to mitigate or avoid such challenges.
\subsection{Technical Challenges}
\subsubsection{Prompt Sensitivity}
The advent of prompt engineering \cite{white2023prompt} has revealed greater potential for LLMs, enabling them to produce outputs better aligned with user needs. However, variations in prompt expression can yield significantly different responses from LLMs \cite{ouyang2024empirical}. This makes code generation particularly unpredictable, especially for more complex tasks. It may be further exacerbated by lengthy prompts, increasing the risk of syntax and structural inconsistencies, commonly referred to as non-determinism. While reducing temperature can mitigate randomness, it can't fully resolve the issue, as no definitive solution currently exists \cite{ouyang2024empirical}.

\textbf{\textit{Why this is a challenge?}} \ \ Research on code generation for both standalone LLMs and LLM-based agents often overlooks reproducibility in actual environment. Benchmarks such as HumanEval, MBPP, and APPS \cite{hendrycks2021measuring} involve short, self-contained Python tasks, requiring only simple prompt strategies such as chain-of-thought \cite{wei2022chain} or few-shot learning \cite{wang2020generalizing}. However, iterative debugging with a maximum iteration limit often encourages researchers to select the best result from multiple attempts, introducing randomness and making replication difficult. Such tendencies result in the ongoing difficulty in achieving reproducibility across studies. 

\textbf{\textit{Suggestion}} \ \ To reduce discrepancies, task description should be specific with added constraints or guidelines to stabilize outputs. In frameworks where LLM-based agents diagnose errors through self-reasoning, the prompt must also be carefully constrained, with few-shot learning to further refine outputs. Additionally, incorporating clarity check from our vision workflow can improve task clarity and reduce hallucinations caused by vague or incomplete prompts.

\subsubsection{Usability \& Consumption}
Usability spans the developer's experience, real-world implementation scenarios, and associated costs. Whether in Q\&A formats (an LLM translating between programming languages) or IDE plugin, LLM-based tools permeate daily programming tasks. However, researches show developers using such plugins do not necessarily see large efficiency gains compared to those who do not \cite{xu2021inidecodegenerationnatural}. This raises a critical question: How can we genuinely enhance the usability of these approaches?

\textbf{\textit{Why this is a challenge?}} \ \ Recent trends show a research shift from pure LLM models to LLM-based agents \cite{jin2024llms}, focusing on applied, macro-level exploration rather than raw model performance. Multi-agent frameworks have recently emerged, harnessing multiple agents for increased flexibility and accuracy \cite{xi2023rise}. While these approaches score impressively on various benchmarks, they often require extensive agent-to-agent interaction, which is time-consuming and struggle to adapt to shifting requirement midstream. Additionally, the `black-box' nature hides internal reasoning and role iteration, reducing trust and comprehension for developers.

Meanwhile, as these frameworks scale, developers may invest more effort orchestrating LLM workflow (crafting prompts, clarifying instructions, building test harnesses) as they would coding manually. Inaccurate output further delay the task with manual debugging. A related concern is whether LLMs proactively request missing information when faced with ambiguities, rather than guessing. Recent community discussions\footnote{\href{https://www.latent.space/p/2024-agents}{https://www.latent.space/p/2024-agents}} emphasize effective agent–human collaboration requires models to recognize knowledge gaps and ask right questions at the right time. Unfortunately, most multi-agent systems still lack robust mechanisms to detect or flag such uncertainties, forcing developers to retroactively diagnose issues. 

\textbf{\textit{Suggestion}} \ \ Studies indicate LLMs perform worse on class-level generation compared to function-level tasks \cite{du2024evaluating}. We argue that future research should prioritize practical usage cases, such as deeper IDE integration and real-world evaluations, rather than only assessing one-off code completions. Token consumption also increases with multi-agent collaborations, as model calls and shared context grows. Effective retrieval-augmented generation and precise content extraction are vital to expand the model's knowledge base without excessive overhead. Moreover, frameworks should bolster the model's ability to detect inaccurate or missing information, promoting clarification or offering alternatives. Developers often avoid LLM-based code generation due to the time spent aligning with and debugging outputs, which negates any efficiency gains. If the model assumes certain functions or libraries exist, usability declines \cite{qian2024chatdev,jiang2024survey}. Our proposed vision workflow (Subsection~\ref{subsec:vision}) addresses these issues by focusing on project-level tasks and adopting a dynamic model cloud. Frequent developer-orchestrator interactions help reduce token cost while improving explainability and code comprehension by revealing the model's assumptions in real time.

\subsubsection{Code Security}
Research on the security of LLM-based code generation has received little attention. Most works focus on bias or jailbreak vulnerabilities in model conversations \cite{duan2024large,deng2023multilingual}, especially for closed-source LLMs, but seldom addresses security concern in code-generation contexts. However, this does not imply that generated code is fully trustworthy or free from hidden security risks in text-based outputs.

\textbf{\textit{Why this is a challenge?}} \ \ In fact, LLM-based code generation can inadvertently introduce security risks by generating unsafe logic or code with latent vulnerabilities \cite{9833571}. It arises not from malicious intent but from the model unintentionally producing insecure code. Furthermore, LLMs often rely on external knowledge sources that may be compromised, exposing outputs to risks like phishing if references (for example, a pip install link) are tampered with \cite{huang2023bias}. The code itself typically does not undergo robust testing; using unverified code (lacking integration tests) in a real development environment that handles sensitive data poses significant risks. Moreover, with LLMs now autonomously calling external APIs and utilizing extensive training data, novel attack surfaces are introduced where backdoor attacks can be placed \cite{zhang2024human}. Multi-agent frameworks further expand the attack surface via user prompts or retrieval data, while current security measures focus on final output, leaving intermediate workflows vulnerable.


\textbf{\textit{Suggestion}} \ \ We suggest future research prioritize code security, together with code functional correctness. A module-level approach can be developed to robust vulnerabilities testing. Furthermore, research has indicated that model-generated code can exhibit biases \cite{huang2024biastestingmitigationllmbased}, highlighting the importance of addressing such issues during the training process.
\subsection{Evaluation Challenges}
Evaluation is one most crucial step in code generation, which ensures that the generated code contains no syntactic or semantic errors and actually fulfills the requirements described in the task.

\textbf{\textit{Why this is a challenge?}} \ \ Currently, there are few widely-used benchmarks for code generation. Examples include HumanEval+ from EvalPlus \cite{liu2024your} and CodeContests, both of which come with test sets ranging from basic function problems to competition level tasks. However, these benchmarks reveal a major limitation: most focus on function-level or single-file tasks, which do not reflect the complexity of real-world software development. In addition, the majority of these benchmarks lack sufficient edge cases, making it difficult to catch hidden bugs or vulnerabilities \cite{liu2024your}. This explains why some studies, despite attaining high pass rates on MBPP or HumanEval, stuggle with more complex problems.

Moreover, evaluation metrics themselves present another challenge. Many existing works rely on simple metrics like Pass@K \cite{zhong2024ldb}, which does not account for efficiency, performance, or maintainability. Consequently, the model may show partial correctness but fail in more realistic settings. In real-world projects, test sets are often too small or nonexistent—some specialized scenarios (e.g., those akin to Alfworld \cite{shridhar2021alfworldaligningtextembodied}) remain untested in code generation tasks, further reducing the scope of usability evaluation.

\textbf{\textit{Suggestion}} \ \ We argue that future research should aim to create more class-level or multi-file benchmarks that better simulate real-world project development. These new tasks should include more complex dependencies and cross-module interactions, enabling us to test how well LLMs handle multi-step integrations instead of trivial function stubs. Existing test sets could be augmented with additional extreme conditions to increase robustness, as demonstrated by EvalPlus \cite{liu2024your}. Beyond simple pass/fail tests, we should also measure the readability and maintainability of the generated code—possibly using agent frameworks to integrate external tools for deeper evaluations. Finally, domain-specific benchmarks (e.g., for financial software) would allow us to reflect real-world demands more accurately. 
\section{Conclusion}
In this paper, we present our vision and reflection for applying large language models in code generation task, examing the technical and evaluation challenges that persist in current research works. We introduce a six-layer collective architecture for how current LLMs researches address code generation task, as well as our vision workflow. Ultimately, we anticipate this work provides a comprehensive landscape of current LLM-based code generation approaches, highlights ongoing critical challenges concerning both technical and evaluation perspectives, and offers practical guidelines for future solutions. 

\clearpage
\balance
\bibliographystyle{ACM-Reference-Format}
\bibliography{sample-base}


\begin{thebibliography}{46}


\ifx \showCODEN    \undefined \def \showCODEN     #1{\unskip}     \fi
\ifx \showISBNx    \undefined \def \showISBNx     #1{\unskip}     \fi
\ifx \showISBNxiii \undefined \def \showISBNxiii  #1{\unskip}     \fi
\ifx \showISSN     \undefined \def \showISSN      #1{\unskip}     \fi
\ifx \showLCCN     \undefined \def \showLCCN      #1{\unskip}     \fi
\ifx \shownote     \undefined \def \shownote      #1{#1}          \fi
\ifx \showarticletitle \undefined \def \showarticletitle #1{#1}   \fi
\ifx \showURL      \undefined \def \showURL       {\relax}        \fi
\providecommand\bibfield[2]{#2}
\providecommand\bibinfo[2]{#2}
\providecommand\natexlab[1]{#1}
\providecommand\showeprint[2][]{arXiv:#2}

\bibitem[Austin et~al\mbox{.}(2021)]%
        {austin2021program}
\bibfield{author}{\bibinfo{person}{Jacob Austin}, \bibinfo{person}{Augustus Odena}, \bibinfo{person}{Maxwell Nye}, \bibinfo{person}{Maarten Bosma}, \bibinfo{person}{Henryk Michalewski}, \bibinfo{person}{David Dohan}, \bibinfo{person}{Ellen Jiang}, \bibinfo{person}{Carrie Cai}, \bibinfo{person}{Michael Terry}, \bibinfo{person}{Quoc Le}, {et~al\mbox{.}}} \bibinfo{year}{2021}\natexlab{}.
\newblock \showarticletitle{Program synthesis with large language models}.
\newblock \bibinfo{journal}{\emph{arXiv preprint arXiv:2108.07732}} (\bibinfo{year}{2021}).
\newblock


\bibitem[Chang et~al\mbox{.}(2024)]%
        {chang2024survey}
\bibfield{author}{\bibinfo{person}{Yupeng Chang}, \bibinfo{person}{Xu Wang}, \bibinfo{person}{Jindong Wang}, \bibinfo{person}{Yuan Wu}, \bibinfo{person}{Linyi Yang}, \bibinfo{person}{Kaijie Zhu}, \bibinfo{person}{Hao Chen}, \bibinfo{person}{Xiaoyuan Yi}, \bibinfo{person}{Cunxiang Wang}, \bibinfo{person}{Yidong Wang}, {et~al\mbox{.}}} \bibinfo{year}{2024}\natexlab{}.
\newblock \showarticletitle{A survey on evaluation of large language models}.
\newblock \bibinfo{journal}{\emph{ACM Transactions on Intelligent Systems and Technology}} \bibinfo{volume}{15}, \bibinfo{number}{3} (\bibinfo{year}{2024}), \bibinfo{pages}{1--45}.
\newblock


\bibitem[Chen et~al\mbox{.}(2021)]%
        {chen2021evaluating}
\bibfield{author}{\bibinfo{person}{Mark Chen}, \bibinfo{person}{Jerry Tworek}, \bibinfo{person}{Heewoo Jun}, \bibinfo{person}{Qiming Yuan}, \bibinfo{person}{Henrique Ponde De~Oliveira Pinto}, \bibinfo{person}{Jared Kaplan}, \bibinfo{person}{Harri Edwards}, \bibinfo{person}{Yuri Burda}, \bibinfo{person}{Nicholas Joseph}, \bibinfo{person}{Greg Brockman}, {et~al\mbox{.}}} \bibinfo{year}{2021}\natexlab{}.
\newblock \showarticletitle{Evaluating large language models trained on code}.
\newblock \bibinfo{journal}{\emph{arXiv preprint arXiv:2107.03374}} (\bibinfo{year}{2021}).
\newblock


\bibitem[Chen et~al\mbox{.}(2023)]%
        {chen2023teaching}
\bibfield{author}{\bibinfo{person}{Xinyun Chen}, \bibinfo{person}{Maxwell Lin}, \bibinfo{person}{Nathanael Sch{\"a}rli}, {and} \bibinfo{person}{Denny Zhou}.} \bibinfo{year}{2023}\natexlab{}.
\newblock \showarticletitle{Teaching large language models to self-debug}.
\newblock \bibinfo{journal}{\emph{arXiv preprint arXiv:2304.05128}} (\bibinfo{year}{2023}).
\newblock


\bibitem[Chen and Chan(2024)]%
        {chen2024large}
\bibfield{author}{\bibinfo{person}{Zenan Chen} {and} \bibinfo{person}{Jason Chan}.} \bibinfo{year}{2024}\natexlab{}.
\newblock \showarticletitle{Large language model in creative work: The role of collaboration modality and user expertise}.
\newblock \bibinfo{journal}{\emph{Management Science}} \bibinfo{volume}{70}, \bibinfo{number}{12} (\bibinfo{year}{2024}), \bibinfo{pages}{9101--9117}.
\newblock


\bibitem[Deng et~al\mbox{.}(2023)]%
        {deng2023multilingual}
\bibfield{author}{\bibinfo{person}{Yue Deng}, \bibinfo{person}{Wenxuan Zhang}, \bibinfo{person}{Sinno~Jialin Pan}, {and} \bibinfo{person}{Lidong Bing}.} \bibinfo{year}{2023}\natexlab{}.
\newblock \showarticletitle{Multilingual jailbreak challenges in large language models}.
\newblock \bibinfo{journal}{\emph{arXiv preprint arXiv:2310.06474}} (\bibinfo{year}{2023}).
\newblock


\bibitem[Du et~al\mbox{.}(2024)]%
        {du2024evaluating}
\bibfield{author}{\bibinfo{person}{Xueying Du}, \bibinfo{person}{Mingwei Liu}, \bibinfo{person}{Kaixin Wang}, \bibinfo{person}{Hanlin Wang}, \bibinfo{person}{Junwei Liu}, \bibinfo{person}{Yixuan Chen}, \bibinfo{person}{Jiayi Feng}, \bibinfo{person}{Chaofeng Sha}, \bibinfo{person}{Xin Peng}, {and} \bibinfo{person}{Yiling Lou}.} \bibinfo{year}{2024}\natexlab{}.
\newblock \showarticletitle{Evaluating large language models in class-level code generation}. In \bibinfo{booktitle}{\emph{Proceedings of the IEEE/ACM 46th International Conference on Software Engineering}}. \bibinfo{pages}{1--13}.
\newblock


\bibitem[Duan(2024)]%
        {duan2024large}
\bibfield{author}{\bibinfo{person}{Yucong Duan}.} \bibinfo{year}{2024}\natexlab{}.
\newblock \showarticletitle{The Large Language Model (LLM) Bias Evaluation (Age Bias)}.
\newblock \bibinfo{journal}{\emph{DIKWP Research Group International Standard Evaluation. DOI}}  \bibinfo{volume}{10} (\bibinfo{year}{2024}).
\newblock


\bibitem[Feng et~al\mbox{.}(2020)]%
        {feng2020codebert}
\bibfield{author}{\bibinfo{person}{Zhangyin Feng}, \bibinfo{person}{Daya Guo}, \bibinfo{person}{Duyu Tang}, \bibinfo{person}{Nan Duan}, \bibinfo{person}{Xiaocheng Feng}, \bibinfo{person}{Ming Gong}, \bibinfo{person}{Linjun Shou}, \bibinfo{person}{Bing Qin}, \bibinfo{person}{Ting Liu}, \bibinfo{person}{Daxin Jiang}, {et~al\mbox{.}}} \bibinfo{year}{2020}\natexlab{}.
\newblock \showarticletitle{Codebert: A pre-trained model for programming and natural languages}.
\newblock \bibinfo{journal}{\emph{arXiv preprint arXiv:2002.08155}} (\bibinfo{year}{2020}).
\newblock


\bibitem[Hadi et~al\mbox{.}(2023)]%
        {hadi2023survey}
\bibfield{author}{\bibinfo{person}{Muhammad~Usman Hadi}, \bibinfo{person}{Rizwan Qureshi}, \bibinfo{person}{Abbas Shah}, \bibinfo{person}{Muhammad Irfan}, \bibinfo{person}{Anas Zafar}, \bibinfo{person}{Muhammad~Bilal Shaikh}, \bibinfo{person}{Naveed Akhtar}, \bibinfo{person}{Jia Wu}, \bibinfo{person}{Seyedali Mirjalili}, {et~al\mbox{.}}} \bibinfo{year}{2023}\natexlab{}.
\newblock \showarticletitle{A survey on large language models: Applications, challenges, limitations, and practical usage}.
\newblock \bibinfo{journal}{\emph{Authorea Preprints}} (\bibinfo{year}{2023}).
\newblock


\bibitem[Hendrycks et~al\mbox{.}(2021)]%
        {hendrycks2021measuring}
\bibfield{author}{\bibinfo{person}{Dan Hendrycks}, \bibinfo{person}{Steven Basart}, \bibinfo{person}{Saurav Kadavath}, \bibinfo{person}{Mantas Mazeika}, \bibinfo{person}{Akul Arora}, \bibinfo{person}{Ethan Guo}, \bibinfo{person}{Collin Burns}, \bibinfo{person}{Samir Puranik}, \bibinfo{person}{Horace He}, \bibinfo{person}{Dawn Song}, {et~al\mbox{.}}} \bibinfo{year}{2021}\natexlab{}.
\newblock \showarticletitle{Measuring coding challenge competence with apps}.
\newblock \bibinfo{journal}{\emph{arXiv preprint arXiv:2105.09938}} (\bibinfo{year}{2021}).
\newblock


\bibitem[Holt et~al\mbox{.}(2023)]%
        {holt2023l2mac}
\bibfield{author}{\bibinfo{person}{Samuel Holt}, \bibinfo{person}{Max~Ruiz Luyten}, {and} \bibinfo{person}{Mihaela van~der Schaar}.} \bibinfo{year}{2023}\natexlab{}.
\newblock \showarticletitle{L2mac: Large language model automatic computer for unbounded code generation}.
\newblock \bibinfo{journal}{\emph{arXiv preprint arXiv:2310.02003}} (\bibinfo{year}{2023}).
\newblock


\bibitem[Hong et~al\mbox{.}(2023)]%
        {hong2023metagpt}
\bibfield{author}{\bibinfo{person}{Sirui Hong}, \bibinfo{person}{Xiawu Zheng}, \bibinfo{person}{Jonathan Chen}, \bibinfo{person}{Yuheng Cheng}, \bibinfo{person}{Jinlin Wang}, \bibinfo{person}{Ceyao Zhang}, \bibinfo{person}{Zili Wang}, \bibinfo{person}{Steven Ka~Shing Yau}, \bibinfo{person}{Zijuan Lin}, \bibinfo{person}{Liyang Zhou}, {et~al\mbox{.}}} \bibinfo{year}{2023}\natexlab{}.
\newblock \showarticletitle{Metagpt: Meta programming for multi-agent collaborative framework}.
\newblock \bibinfo{journal}{\emph{arXiv preprint arXiv:2308.00352}} (\bibinfo{year}{2023}).
\newblock


\bibitem[Huang et~al\mbox{.}(2023a)]%
        {huang2023codecot}
\bibfield{author}{\bibinfo{person}{Dong Huang}, \bibinfo{person}{Qingwen Bu}, {and} \bibinfo{person}{Heming Cui}.} \bibinfo{year}{2023}\natexlab{a}.
\newblock \showarticletitle{Codecot and beyond: Learning to program and test like a developer}.
\newblock \bibinfo{journal}{\emph{arXiv preprint arXiv:2308.08784}} (\bibinfo{year}{2023}).
\newblock


\bibitem[Huang et~al\mbox{.}(2023b)]%
        {huang2023bias}
\bibfield{author}{\bibinfo{person}{Dong Huang}, \bibinfo{person}{Qingwen Bu}, \bibinfo{person}{Jie Zhang}, \bibinfo{person}{Xiaofei Xie}, \bibinfo{person}{Junjie Chen}, {and} \bibinfo{person}{Heming Cui}.} \bibinfo{year}{2023}\natexlab{b}.
\newblock \showarticletitle{Bias assessment and mitigation in llm-based code generation}.
\newblock \bibinfo{journal}{\emph{arXiv preprint arXiv:2309.14345}} (\bibinfo{year}{2023}).
\newblock


\bibitem[Huang et~al\mbox{.}(2024)]%
        {huang2024biastestingmitigationllmbased}
\bibfield{author}{\bibinfo{person}{Dong Huang}, \bibinfo{person}{Qingwen Bu}, \bibinfo{person}{Jie Zhang}, \bibinfo{person}{Xiaofei Xie}, \bibinfo{person}{Junjie Chen}, {and} \bibinfo{person}{Heming Cui}.} \bibinfo{year}{2024}\natexlab{}.
\newblock \bibinfo{title}{Bias Testing and Mitigation in LLM-based Code Generation}.
\newblock
\showeprint[arxiv]{2309.14345}~[cs.SE]
\urldef\tempurl%
\url{https://arxiv.org/abs/2309.14345}
\showURL{%
\tempurl}


\bibitem[Huang et~al\mbox{.}(2023c)]%
        {huang2023chatgpt}
\bibfield{author}{\bibinfo{person}{Hanyao Huang}, \bibinfo{person}{Ou Zheng}, \bibinfo{person}{Dongdong Wang}, \bibinfo{person}{Jiayi Yin}, \bibinfo{person}{Zijin Wang}, \bibinfo{person}{Shengxuan Ding}, \bibinfo{person}{Heng Yin}, \bibinfo{person}{Chuan Xu}, \bibinfo{person}{Renjie Yang}, \bibinfo{person}{Qian Zheng}, {et~al\mbox{.}}} \bibinfo{year}{2023}\natexlab{c}.
\newblock \showarticletitle{ChatGPT for shaping the future of dentistry: the potential of multi-modal large language model}.
\newblock \bibinfo{journal}{\emph{International Journal of Oral Science}} \bibinfo{volume}{15}, \bibinfo{number}{1} (\bibinfo{year}{2023}), \bibinfo{pages}{29}.
\newblock


\bibitem[Ji et~al\mbox{.}(2023)]%
        {ji2023survey}
\bibfield{author}{\bibinfo{person}{Ziwei Ji}, \bibinfo{person}{Nayeon Lee}, \bibinfo{person}{Rita Frieske}, \bibinfo{person}{Tiezheng Yu}, \bibinfo{person}{Dan Su}, \bibinfo{person}{Yan Xu}, \bibinfo{person}{Etsuko Ishii}, \bibinfo{person}{Ye~Jin Bang}, \bibinfo{person}{Andrea Madotto}, {and} \bibinfo{person}{Pascale Fung}.} \bibinfo{year}{2023}\natexlab{}.
\newblock \showarticletitle{Survey of hallucination in natural language generation}.
\newblock \bibinfo{journal}{\emph{Comput. Surveys}} \bibinfo{volume}{55}, \bibinfo{number}{12} (\bibinfo{year}{2023}), \bibinfo{pages}{1--38}.
\newblock


\bibitem[Jiang et~al\mbox{.}(2024)]%
        {jiang2024survey}
\bibfield{author}{\bibinfo{person}{Juyong Jiang}, \bibinfo{person}{Fan Wang}, \bibinfo{person}{Jiasi Shen}, \bibinfo{person}{Sungju Kim}, {and} \bibinfo{person}{Sunghun Kim}.} \bibinfo{year}{2024}\natexlab{}.
\newblock \showarticletitle{A Survey on Large Language Models for Code Generation}.
\newblock \bibinfo{journal}{\emph{arXiv preprint arXiv:2406.00515}} (\bibinfo{year}{2024}).
\newblock


\bibitem[Jin et~al\mbox{.}(2024)]%
        {jin2024llms}
\bibfield{author}{\bibinfo{person}{Haolin Jin}, \bibinfo{person}{Linghan Huang}, \bibinfo{person}{Haipeng Cai}, \bibinfo{person}{Jun Yan}, \bibinfo{person}{Bo Li}, {and} \bibinfo{person}{Huaming Chen}.} \bibinfo{year}{2024}\natexlab{}.
\newblock \showarticletitle{From llms to llm-based agents for software engineering: A survey of current, challenges and future}.
\newblock \bibinfo{journal}{\emph{arXiv preprint arXiv:2408.02479}} (\bibinfo{year}{2024}).
\newblock


\bibitem[Kaddour et~al\mbox{.}(2023)]%
        {kaddour2023challenges}
\bibfield{author}{\bibinfo{person}{Jean Kaddour}, \bibinfo{person}{Joshua Harris}, \bibinfo{person}{Maximilian Mozes}, \bibinfo{person}{Herbie Bradley}, \bibinfo{person}{Roberta Raileanu}, {and} \bibinfo{person}{Robert McHardy}.} \bibinfo{year}{2023}\natexlab{}.
\newblock \showarticletitle{Challenges and applications of large language models}.
\newblock \bibinfo{journal}{\emph{arXiv preprint arXiv:2307.10169}} (\bibinfo{year}{2023}).
\newblock


\bibitem[Kalyan(2024)]%
        {kalyan2024survey}
\bibfield{author}{\bibinfo{person}{Katikapalli~Subramanyam Kalyan}.} \bibinfo{year}{2024}\natexlab{}.
\newblock \showarticletitle{A survey of GPT-3 family large language models including ChatGPT and GPT-4}.
\newblock \bibinfo{journal}{\emph{Natural Language Processing Journal}}  \bibinfo{volume}{6} (\bibinfo{year}{2024}), \bibinfo{pages}{100048}.
\newblock


\bibitem[Lewis et~al\mbox{.}(2020)]%
        {lewis2020retrieval}
\bibfield{author}{\bibinfo{person}{Patrick Lewis}, \bibinfo{person}{Ethan Perez}, \bibinfo{person}{Aleksandra Piktus}, \bibinfo{person}{Fabio Petroni}, \bibinfo{person}{Vladimir Karpukhin}, \bibinfo{person}{Naman Goyal}, \bibinfo{person}{Heinrich K{\"u}ttler}, \bibinfo{person}{Mike Lewis}, \bibinfo{person}{Wen-tau Yih}, \bibinfo{person}{Tim Rockt{\"a}schel}, {et~al\mbox{.}}} \bibinfo{year}{2020}\natexlab{}.
\newblock \showarticletitle{Retrieval-augmented generation for knowledge-intensive nlp tasks}.
\newblock \bibinfo{journal}{\emph{Advances in Neural Information Processing Systems}}  \bibinfo{volume}{33} (\bibinfo{year}{2020}), \bibinfo{pages}{9459--9474}.
\newblock


\bibitem[Lin et~al\mbox{.}(2024)]%
        {lin2024soen}
\bibfield{author}{\bibinfo{person}{Feng Lin}, \bibinfo{person}{Dong~Jae Kim}, {and} \bibinfo{person}{TH Chen}.} \bibinfo{year}{2024}\natexlab{}.
\newblock \showarticletitle{SOEN-101: Code Generation by Emulating Software Process Models Using Large Language Model Agents}.
\newblock \bibinfo{journal}{\emph{arXiv preprint arXiv:2403.15852}} (\bibinfo{year}{2024}).
\newblock


\bibitem[Liu et~al\mbox{.}(2024)]%
        {liu2024your}
\bibfield{author}{\bibinfo{person}{Jiawei Liu}, \bibinfo{person}{Chunqiu~Steven Xia}, \bibinfo{person}{Yuyao Wang}, {and} \bibinfo{person}{Lingming Zhang}.} \bibinfo{year}{2024}\natexlab{}.
\newblock \showarticletitle{Is your code generated by chatgpt really correct? rigorous evaluation of large language models for code generation}.
\newblock \bibinfo{journal}{\emph{Advances in Neural Information Processing Systems}}  \bibinfo{volume}{36} (\bibinfo{year}{2024}).
\newblock


\bibitem[Lu et~al\mbox{.}(2021)]%
        {lu2021codexglue}
\bibfield{author}{\bibinfo{person}{Shuai Lu}, \bibinfo{person}{Daya Guo}, \bibinfo{person}{Shuo Ren}, \bibinfo{person}{Junjie Huang}, \bibinfo{person}{Alexey Svyatkovskiy}, \bibinfo{person}{Ambrosio Blanco}, \bibinfo{person}{Colin Clement}, \bibinfo{person}{Dawn Drain}, \bibinfo{person}{Daxin Jiang}, \bibinfo{person}{Duyu Tang}, {et~al\mbox{.}}} \bibinfo{year}{2021}\natexlab{}.
\newblock \showarticletitle{Codexglue: A machine learning benchmark dataset for code understanding and generation}.
\newblock \bibinfo{journal}{\emph{arXiv preprint arXiv:2102.04664}} (\bibinfo{year}{2021}).
\newblock


\bibitem[Mu et~al\mbox{.}(2023)]%
        {mu2023clarifygpt}
\bibfield{author}{\bibinfo{person}{Fangwen Mu}, \bibinfo{person}{Lin Shi}, \bibinfo{person}{Song Wang}, \bibinfo{person}{Zhuohao Yu}, \bibinfo{person}{Binquan Zhang}, \bibinfo{person}{Chenxue Wang}, \bibinfo{person}{Shichao Liu}, {and} \bibinfo{person}{Qing Wang}.} \bibinfo{year}{2023}\natexlab{}.
\newblock \showarticletitle{ClarifyGPT: Empowering LLM-based Code Generation with Intention Clarification}.
\newblock \bibinfo{journal}{\emph{arXiv preprint arXiv:2310.10996}} (\bibinfo{year}{2023}).
\newblock


\bibitem[Ni et~al\mbox{.}(2024)]%
        {ni2024l2ceval}
\bibfield{author}{\bibinfo{person}{Ansong Ni}, \bibinfo{person}{Pengcheng Yin}, \bibinfo{person}{Yilun Zhao}, \bibinfo{person}{Martin Riddell}, \bibinfo{person}{Troy Feng}, \bibinfo{person}{Rui Shen}, \bibinfo{person}{Stephen Yin}, \bibinfo{person}{Ye Liu}, \bibinfo{person}{Semih Yavuz}, \bibinfo{person}{Caiming Xiong}, {et~al\mbox{.}}} \bibinfo{year}{2024}\natexlab{}.
\newblock \showarticletitle{L2ceval: Evaluating language-to-code generation capabilities of large language models}.
\newblock \bibinfo{journal}{\emph{Transactions of the Association for Computational Linguistics}}  \bibinfo{volume}{12} (\bibinfo{year}{2024}), \bibinfo{pages}{1311--1329}.
\newblock


\bibitem[Nijkamp et~al\mbox{.}(2022)]%
        {nijkamp2022codegen}
\bibfield{author}{\bibinfo{person}{Erik Nijkamp}, \bibinfo{person}{Bo Pang}, \bibinfo{person}{Hiroaki Hayashi}, \bibinfo{person}{Lifu Tu}, \bibinfo{person}{Huan Wang}, \bibinfo{person}{Yingbo Zhou}, \bibinfo{person}{Silvio Savarese}, {and} \bibinfo{person}{Caiming Xiong}.} \bibinfo{year}{2022}\natexlab{}.
\newblock \showarticletitle{Codegen: An open large language model for code with multi-turn program synthesis}.
\newblock \bibinfo{journal}{\emph{arXiv preprint arXiv:2203.13474}} (\bibinfo{year}{2022}).
\newblock


\bibitem[Ouyang et~al\mbox{.}(2024)]%
        {ouyang2024empirical}
\bibfield{author}{\bibinfo{person}{Shuyin Ouyang}, \bibinfo{person}{Jie~M Zhang}, \bibinfo{person}{Mark Harman}, {and} \bibinfo{person}{Meng Wang}.} \bibinfo{year}{2024}\natexlab{}.
\newblock \showarticletitle{An empirical study of the non-determinism of chatgpt in code generation}.
\newblock \bibinfo{journal}{\emph{ACM Transactions on Software Engineering and Methodology}} (\bibinfo{year}{2024}).
\newblock


\bibitem[Pearce et~al\mbox{.}(2022)]%
        {9833571}
\bibfield{author}{\bibinfo{person}{Hammond Pearce}, \bibinfo{person}{Baleegh Ahmad}, \bibinfo{person}{Benjamin Tan}, \bibinfo{person}{Brendan Dolan-Gavitt}, {and} \bibinfo{person}{Ramesh Karri}.} \bibinfo{year}{2022}\natexlab{}.
\newblock \showarticletitle{Asleep at the Keyboard? Assessing the Security of GitHub Copilot’s Code Contributions}. In \bibinfo{booktitle}{\emph{2022 IEEE Symposium on Security and Privacy (SP)}}. \bibinfo{pages}{754--768}.
\newblock
\href{https://doi.org/10.1109/SP46214.2022.9833571}{doi:\nolinkurl{10.1109/SP46214.2022.9833571}}


\bibitem[Qian et~al\mbox{.}(2024)]%
        {qian2024chatdev}
\bibfield{author}{\bibinfo{person}{Chen Qian}, \bibinfo{person}{Wei Liu}, \bibinfo{person}{Hongzhang Liu}, \bibinfo{person}{Nuo Chen}, \bibinfo{person}{Yufan Dang}, \bibinfo{person}{Jiahao Li}, \bibinfo{person}{Cheng Yang}, \bibinfo{person}{Weize Chen}, \bibinfo{person}{Yusheng Su}, \bibinfo{person}{Xin Cong}, {et~al\mbox{.}}} \bibinfo{year}{2024}\natexlab{}.
\newblock \showarticletitle{Chatdev: Communicative agents for software development}. In \bibinfo{booktitle}{\emph{Proceedings of the 62nd Annual Meeting of the Association for Computational Linguistics (Volume 1: Long Papers)}}. \bibinfo{pages}{15174--15186}.
\newblock


\bibitem[Ruder et~al\mbox{.}(2019)]%
        {ruder2019unsupervised}
\bibfield{author}{\bibinfo{person}{Sebastian Ruder}, \bibinfo{person}{Anders S{\o}gaard}, {and} \bibinfo{person}{Ivan Vuli{\'c}}.} \bibinfo{year}{2019}\natexlab{}.
\newblock \showarticletitle{Unsupervised cross-lingual representation learning}. In \bibinfo{booktitle}{\emph{Proceedings of the 57th Annual Meeting of the Association for Computational Linguistics: Tutorial Abstracts}}. \bibinfo{pages}{31--38}.
\newblock


\bibitem[Shridhar et~al\mbox{.}(2021)]%
        {shridhar2021alfworldaligningtextembodied}
\bibfield{author}{\bibinfo{person}{Mohit Shridhar}, \bibinfo{person}{Xingdi Yuan}, \bibinfo{person}{Marc-Alexandre Côté}, \bibinfo{person}{Yonatan Bisk}, \bibinfo{person}{Adam Trischler}, {and} \bibinfo{person}{Matthew Hausknecht}.} \bibinfo{year}{2021}\natexlab{}.
\newblock \bibinfo{title}{ALFWorld: Aligning Text and Embodied Environments for Interactive Learning}.
\newblock
\showeprint[arxiv]{2010.03768}~[cs.CL]
\urldef\tempurl%
\url{https://arxiv.org/abs/2010.03768}
\showURL{%
\tempurl}


\bibitem[Wang et~al\mbox{.}(2020)]%
        {wang2020generalizing}
\bibfield{author}{\bibinfo{person}{Yaqing Wang}, \bibinfo{person}{Quanming Yao}, \bibinfo{person}{James~T Kwok}, {and} \bibinfo{person}{Lionel~M Ni}.} \bibinfo{year}{2020}\natexlab{}.
\newblock \showarticletitle{Generalizing from a few examples: A survey on few-shot learning}.
\newblock \bibinfo{journal}{\emph{ACM computing surveys (csur)}} \bibinfo{volume}{53}, \bibinfo{number}{3} (\bibinfo{year}{2020}), \bibinfo{pages}{1--34}.
\newblock


\bibitem[Weerasiri et~al\mbox{.}(2017)]%
        {weerasiri2017taxonomy}
\bibfield{author}{\bibinfo{person}{Denis Weerasiri}, \bibinfo{person}{Moshe~Chai Barukh}, \bibinfo{person}{Boualem Benatallah}, \bibinfo{person}{Quan~Z Sheng}, {and} \bibinfo{person}{Rajiv Ranjan}.} \bibinfo{year}{2017}\natexlab{}.
\newblock \showarticletitle{A taxonomy and survey of cloud resource orchestration techniques}.
\newblock \bibinfo{journal}{\emph{ACM Computing Surveys (CSUR)}} \bibinfo{volume}{50}, \bibinfo{number}{2} (\bibinfo{year}{2017}), \bibinfo{pages}{1--41}.
\newblock


\bibitem[Wei et~al\mbox{.}(2022)]%
        {wei2022chain}
\bibfield{author}{\bibinfo{person}{Jason Wei}, \bibinfo{person}{Xuezhi Wang}, \bibinfo{person}{Dale Schuurmans}, \bibinfo{person}{Maarten Bosma}, \bibinfo{person}{Fei Xia}, \bibinfo{person}{Ed Chi}, \bibinfo{person}{Quoc~V Le}, \bibinfo{person}{Denny Zhou}, {et~al\mbox{.}}} \bibinfo{year}{2022}\natexlab{}.
\newblock \showarticletitle{Chain-of-thought prompting elicits reasoning in large language models}.
\newblock \bibinfo{journal}{\emph{Advances in neural information processing systems}}  \bibinfo{volume}{35} (\bibinfo{year}{2022}), \bibinfo{pages}{24824--24837}.
\newblock


\bibitem[White et~al\mbox{.}(2023)]%
        {white2023prompt}
\bibfield{author}{\bibinfo{person}{Jules White}, \bibinfo{person}{Quchen Fu}, \bibinfo{person}{Sam Hays}, \bibinfo{person}{Michael Sandborn}, \bibinfo{person}{Carlos Olea}, \bibinfo{person}{Henry Gilbert}, \bibinfo{person}{Ashraf Elnashar}, \bibinfo{person}{Jesse Spencer-Smith}, {and} \bibinfo{person}{Douglas~C Schmidt}.} \bibinfo{year}{2023}\natexlab{}.
\newblock \showarticletitle{A prompt pattern catalog to enhance prompt engineering with chatgpt}.
\newblock \bibinfo{journal}{\emph{arXiv preprint arXiv:2302.11382}} (\bibinfo{year}{2023}).
\newblock


\bibitem[Wu et~al\mbox{.}(2023)]%
        {wu2023next}
\bibfield{author}{\bibinfo{person}{Shengqiong Wu}, \bibinfo{person}{Hao Fei}, \bibinfo{person}{Leigang Qu}, \bibinfo{person}{Wei Ji}, {and} \bibinfo{person}{Tat-Seng Chua}.} \bibinfo{year}{2023}\natexlab{}.
\newblock \showarticletitle{Next-gpt: Any-to-any multimodal llm}.
\newblock \bibinfo{journal}{\emph{arXiv preprint arXiv:2309.05519}} (\bibinfo{year}{2023}).
\newblock


\bibitem[Xi et~al\mbox{.}(2023)]%
        {xi2023rise}
\bibfield{author}{\bibinfo{person}{Zhiheng Xi}, \bibinfo{person}{Wenxiang Chen}, \bibinfo{person}{Xin Guo}, \bibinfo{person}{Wei He}, \bibinfo{person}{Yiwen Ding}, \bibinfo{person}{Boyang Hong}, \bibinfo{person}{Ming Zhang}, \bibinfo{person}{Junzhe Wang}, \bibinfo{person}{Senjie Jin}, \bibinfo{person}{Enyu Zhou}, {et~al\mbox{.}}} \bibinfo{year}{2023}\natexlab{}.
\newblock \showarticletitle{The rise and potential of large language model based agents: A survey}.
\newblock \bibinfo{journal}{\emph{arXiv preprint arXiv:2309.07864}} (\bibinfo{year}{2023}).
\newblock


\bibitem[Xu et~al\mbox{.}(2021)]%
        {xu2021inidecodegenerationnatural}
\bibfield{author}{\bibinfo{person}{Frank~F. Xu}, \bibinfo{person}{Bogdan Vasilescu}, {and} \bibinfo{person}{Graham Neubig}.} \bibinfo{year}{2021}\natexlab{}.
\newblock \bibinfo{title}{In-IDE Code Generation from Natural Language: Promise and Challenges}.
\newblock
\showeprint[arxiv]{2101.11149}~[cs.SE]
\urldef\tempurl%
\url{https://arxiv.org/abs/2101.11149}
\showURL{%
\tempurl}


\bibitem[Yang et~al\mbox{.}(2024)]%
        {yang2024robustnesssecurityprivacyexplainability}
\bibfield{author}{\bibinfo{person}{Zhou Yang}, \bibinfo{person}{Zhensu Sun}, \bibinfo{person}{Terry~Zhuo Yue}, \bibinfo{person}{Premkumar Devanbu}, {and} \bibinfo{person}{David Lo}.} \bibinfo{year}{2024}\natexlab{}.
\newblock \bibinfo{title}{Robustness, Security, Privacy, Explainability, Efficiency, and Usability of Large Language Models for Code}.
\newblock
\showeprint[arxiv]{2403.07506}~[cs.SE]
\urldef\tempurl%
\url{https://arxiv.org/abs/2403.07506}
\showURL{%
\tempurl}


\bibitem[Zhang et~al\mbox{.}(2023)]%
        {zhang2023self}
\bibfield{author}{\bibinfo{person}{Kechi Zhang}, \bibinfo{person}{Zhuo Li}, \bibinfo{person}{Jia Li}, \bibinfo{person}{Ge Li}, {and} \bibinfo{person}{Zhi Jin}.} \bibinfo{year}{2023}\natexlab{}.
\newblock \showarticletitle{Self-edit: Fault-aware code editor for code generation}.
\newblock \bibinfo{journal}{\emph{arXiv preprint arXiv:2305.04087}} (\bibinfo{year}{2023}).
\newblock


\bibitem[Zhang et~al\mbox{.}(2024b)]%
        {zhang2024human}
\bibfield{author}{\bibinfo{person}{Quan Zhang}, \bibinfo{person}{Binqi Zeng}, \bibinfo{person}{Chijin Zhou}, \bibinfo{person}{Gwihwan Go}, \bibinfo{person}{Heyuan Shi}, {and} \bibinfo{person}{Yu Jiang}.} \bibinfo{year}{2024}\natexlab{b}.
\newblock \showarticletitle{Human-imperceptible retrieval poisoning attacks in LLM-powered applications}. In \bibinfo{booktitle}{\emph{Companion Proceedings of the 32nd ACM International Conference on the Foundations of Software Engineering}}. \bibinfo{pages}{502--506}.
\newblock


\bibitem[Zhang et~al\mbox{.}(2024a)]%
        {zhang2024experimenting}
\bibfield{author}{\bibinfo{person}{Simiao Zhang}, \bibinfo{person}{Jiaping Wang}, \bibinfo{person}{Guoliang Dong}, \bibinfo{person}{Jun Sun}, \bibinfo{person}{Yueling Zhang}, {and} \bibinfo{person}{Geguang Pu}.} \bibinfo{year}{2024}\natexlab{a}.
\newblock \showarticletitle{Experimenting a New Programming Practice with LLMs}.
\newblock \bibinfo{journal}{\emph{arXiv preprint arXiv:2401.01062}} (\bibinfo{year}{2024}).
\newblock


\bibitem[Zhong et~al\mbox{.}(2024)]%
        {zhong2024ldb}
\bibfield{author}{\bibinfo{person}{Li Zhong}, \bibinfo{person}{Zilong Wang}, {and} \bibinfo{person}{Jingbo Shang}.} \bibinfo{year}{2024}\natexlab{}.
\newblock \showarticletitle{Ldb: A large language model debugger via verifying runtime execution step-by-step}.
\newblock \bibinfo{journal}{\emph{arXiv preprint arXiv:2402.16906}} (\bibinfo{year}{2024}).
\newblock


\end{thebibliography}

\end{document}